\documentclass[12pt,preprint]{aastex}

\shorttitle{Pulsations of $\beta$ CMi}
\shortauthors{}

\begin{document}

\title{{\it MOST}\footnotemark[1]~ DETECTS g-MODES IN THE LATE-TYPE Be STAR \\
$\beta$ CMi (B8Ve)}

\footnotetext[1]{Based on data from the {\it MOST} satellite, a Canadian Space Agency mission, jointly operated by Dynacon Inc., the University of Toronto Institute of
Aerospace Studies and the University of British Columbia with the assistance of the University of Vienna.} 

\author{H. Saio\altaffilmark{2},
  C. Cameron\altaffilmark{3}, R. Kuschnig\altaffilmark{3},
G.A.H. Walker\altaffilmark{4}, J.M. Matthews\altaffilmark{3},  J.F. Rowe\altaffilmark{3},
U. Lee\altaffilmark{2}, D. Huber\altaffilmark{5}, W.W. Weiss\altaffilmark{5},
D.B. Guenther\altaffilmark{6}, A.F.J. Moffat\altaffilmark{7}, 
S.M. Rucinski\altaffilmark{8}, D. Sasselov\altaffilmark{9} 
} 

\altaffiltext{2}{Astronomical Institute, Graduate School of Science, Tohoku University, Sendai, 980-8578, Japan;
saio@astr.tohuku.ac.jp, lee@astr.tohoku.ac.jp}

\altaffiltext{3}{Dept. Physics and Astronomy, University of British Columbia, 
6224 Agricultural Road, Vancouver, BC V6T 1Z1, Canada;  kuschnig@astro.phys.ubc.ca, matthews@phas.ubc.ca, ccameron@phas.ubc.ca, rowe@phas.ubc.ca}

\altaffiltext{4}{1234 Hewlett Place, Victoria, BC V8S 4P7, Canada;
gordonwa@uvic.ca}

\altaffiltext{5}{Institut f\"ur Astronomie, Universit\"at Wien 
T\"urkenschanzstrasse 17, A--1180 Wien, Austria;
a0206892@unet.univie.ac.at, weiss@astro.univie.ac.at}

\altaffiltext{6}{Department of Astronomy and Physics, St. Mary's University
Halifax, NS B3H 3C3, Canada;
guenther@ap.stmarys.ca}

\altaffiltext{7}{D\'ept. de physique, Univ. de Montr\'eal 
C.P.\ 6128, Succ.\ Centre-Ville, Montr\'eal, QC H3C 3J7, Canada;
and Obs. du mont M\'egantic;
moffat@astro.umontreal.ca}

\altaffiltext{8}{Dept. Astronomy \& Astrophysics, David Dunlap Obs., Univ. Toronto 
P.O.~Box 360, Richmond Hill, ON L4C 4Y6, Canada;
rucinski@astro.utoronto.ca}

\altaffiltext{9}{Harvard-Smithsonian Center for Astrophysics, 
60 Garden Street, Cambridge, MA 02138, USA;
sasselov@cfa.harvard.edu}

\begin{abstract}
The {\it Microvariability and Oscillations of stars (MOST)} satellite has
detected low-amplitude light variations ($\Delta m\sim$1 mmag) in the Be star 
$\beta$ CMi (B8Ve). The observations lasted 41 days and the variations have 
typical periods $\sim 0.3$ days. We demonstrate that the dominant frequencies are 
consistent with prograde high-order g-modes of $m=-1$ excited by the Fe-bump 
of opacity in an intermediate-mass ($\approx 3.5 M_\odot$) star with a nearly 
critical rotation period of 0.38 days. This is the first detection of nonradial 
g-mode pulsations in a Be star later than B6 leading to the possibility that 
pulsations are excited in all classical Be stars.
\end{abstract}


\section{INTRODUCTION}
Be stars are non-supergiant, rapid rotators, which at times show 
emission in the Balmer and certain metallic lines.
They are often called ``classical Be stars'' to distinguish them from other  
emission line stars such as Herbig Ae/Be stars and Algol systems
\citep[see][for a recent review on Be stars]{Por03}. 
The line emission is assumed to arise in a geometrically 
thin disk consisting of matter ejected from the central star. 
Such circumstellar disks have actually been imaged by long-baseline
interferometry for several Be stars including $\beta$ CMi \citep[e.g.][]{Tyc05}. 
Our target star, $\beta$ CMi (HD 58715, HR 2845), is 
a bright but 'quiet' late-type Be star (B8Ve; $V=2.886, B-V=-0.072$);
no photometric variations were found by \citet{Pav97}. 
The star shows only gradual and small long-term variations 
in  H$\alpha$ emission \citep[e.g.,][]{Han96} arising from a relatively
small disk of $\sim 3$ stellar radii \citep[e.g.][]{Tyc05}.

Other Be stars, particularly of early-type, often show short-term photometric 
variations of less than a few days. The causes of these variations are assumed 
to be pulsation and rotational modulation \citep[see][for a review]{Por03}. 
Many early-type Be stars exhibit line-profile variations indicating the presence 
of non-radial pulsations  \citep[e.g.][]{Riv03}.
Multi-periodic photometric variations are also the signature of non-radial 
pulsations such as those found recently by {\it MOST} for $\zeta$ Oph 
\citep[O9.5Ve;][]{Wal05a} and HD 163868 \citep[B5Ve;][]{Wal05b}. Until now, 
there has been no clear indication of pulsations for late-type Be stars. 
\citet{Baa89} searched for, but failed to detect any, line-profile variations 
in B8-B9.5 stars. There were very small photometric variations in a few 
late-type Be stars, but these were judged to be mono-periodic 
(e.g., $\mu$ Pic B9Ve, Balona et al 1992; $\zeta$ Crv B8Vne, Barrera et al. 1991)
and attributed to rotational modulation.

It is of critical importance to know if nonradial pulsations are excited even 
in late-type Be stars because it would indicate that 
all Be stars may have nonradial pulsations which
could play a critical role in mass ejection. 
Excitation of g-modes in late-type Be stars is likely because
the cool boundary of the distribution of Be stars on the HR diagram 
\citep{Zor05} roughly coincides with the cool boundary of the 
Slowly Pulsating B (SPB) 
stars \citep{Pam99}, and because prograde g-modes can be excited
even in a rapidly rotating star \citep{Wal05b}. 
With this motivation we observed $\beta$ CMi with {\it MOST} and indeed discovered 
multi-periodic light variations, albeit of low amplitude. 

\section{ THE {\it MOST} PHOTOMETRY}
\label{most}

The {\it MOST} satellite was launched on June 2003 and the original mission is 
described by \citet{MOST}. 
A 15/17.3 cm Rumak-Maksutov telescope
feeds two CCDs, one for tracking and the other for science, through a single, 
custom, broadband filter (350 -- 700 nm). Starlight from 
Primary Science Targets ($V \leq 6$) 
is projected onto the science CCD as a fixed (Fabry) image of the telescope 
pupil covering about 1500 pixels for high photometric stability and 
insensitivity to detector flatfield irregularities and the effect of particle 
irradiation on individual pixels. The experiment was designed to detect 
photometric variations with periods of minutes at micro-magnitude precision 
and does not rely on comparison stars or flat-fielding for the Fabry photometry. 
There is no direct connection to any photometric system. Tracking jitter was 
dramatically reduced by early 2004 to $\sim$1 arcsec which led to significantly 
higher precision in the Fabry photometry.

The observations received from the satellite were reduced by RK. Outlying data 
points generated by poor tracking or cosmic ray hits were removed. 
{\it MOST} suffers from parasitic light, mostly Earthshine, at certain 
orbital phases, with the amount and phase depending on the stellar coordinates, 
spacecraft roll and season of the year. Data are also recorded for Fabry images 
from seven of the eight lenses adjacent to the target Fabry lens in order to 
track the stray light background. These background signals were combined in 
a mean and subtracted from the target photometry. This also corrected for bias, 
dark and background signals and their variations. The reductions basically 
followed the scheme outlined earlier by \citet{Ruc04}.

{\it MOST} observed $\beta$ CMi from December 16 2005 till  
January 26 2006 for a total of 41 days. Observing
time was split between $\beta$ CMi and another field
such that only 45\% of each 101 min orbit 
was available for $\beta$ CMi. In addition to the 45\% duty cycle 
over the 41 days, there was a single $\sim$12 hour gap. 
Individual exposures were 8 sec taken every 
20 sec (the sampling time).  For the final frequency
analysis mean magnitudes were calculated from data accumulated every 5 min.
The RMS scatter of the individual 8-sec exposures is 0.300 mmag 
and 0.070 mmag for the the 5-min means. 

\begin{figure}
\epsscale{0.8}
\plotone{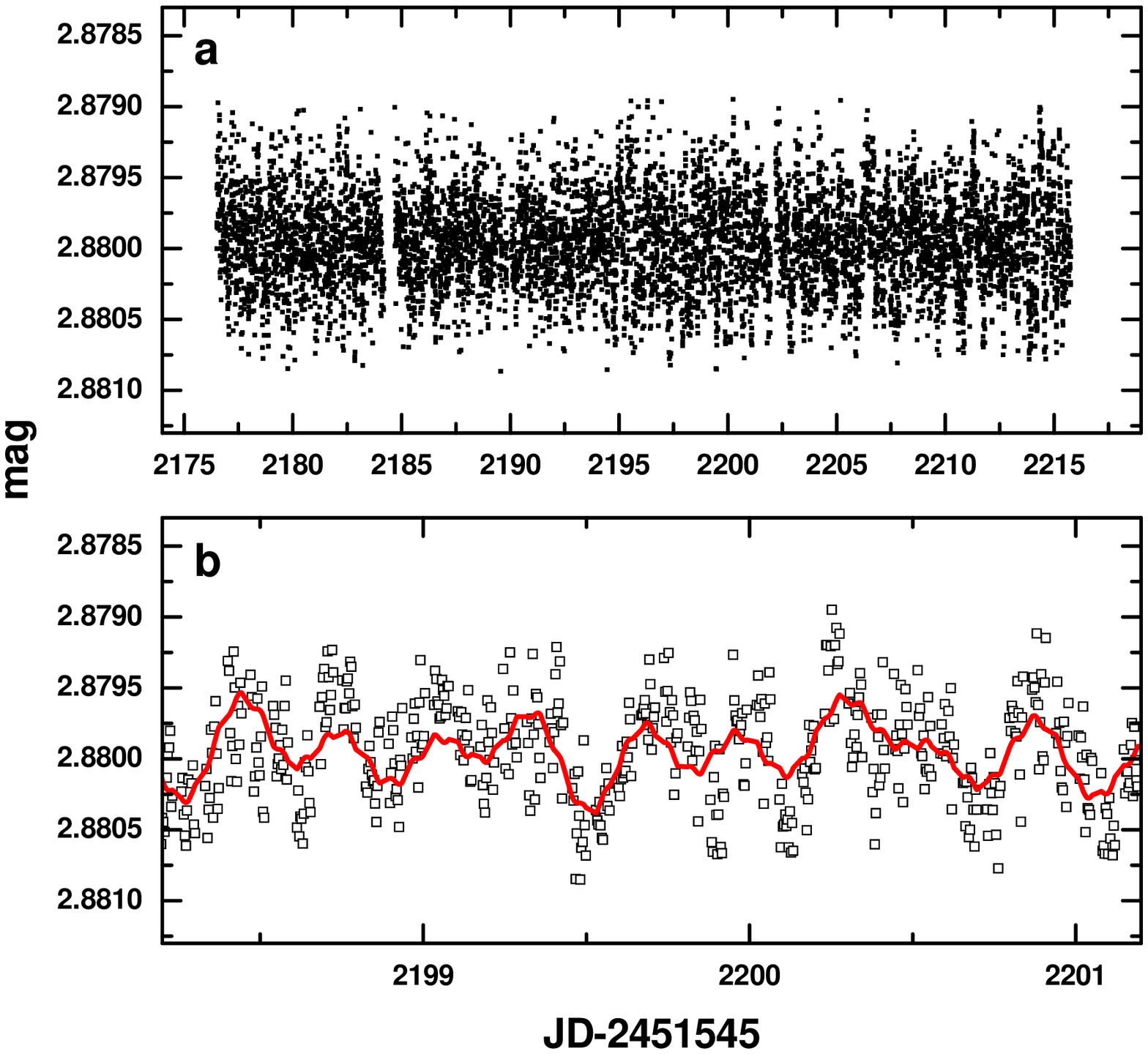}
\caption{a) {\it MOST} 41-day light curve of $\beta$ CMi. Each point is 
a 5-min mean and the point-to-point RMS scatter is 0.070 mmag. 
b) An expanded three day portion of the light curve in (a). 
The solid line is the fit of the 20 highest peaks from the frequency analysis 
of the full light curve.}
\label{obs}
\end{figure}

Figure~\ref{obs} displays the full 41-day light curve for the 5-min means with 
time in heliocentric JD. A 3-day expanded portion of the light curve is shown 
in the second panel.  The solid line in the lower panel is 
the reconstructed lightcurve obtained with 
the most significant peaks in the frequency analysis discussed below. 
The complete light curve can be downloaded from the MOST Public Data Archive
\footnote{The {\it MOST} Public Data Archive is available 
at www.astro.ubc.ca/MOST.}.

\begin{figure}
\epsscale{0.7}
\plotone{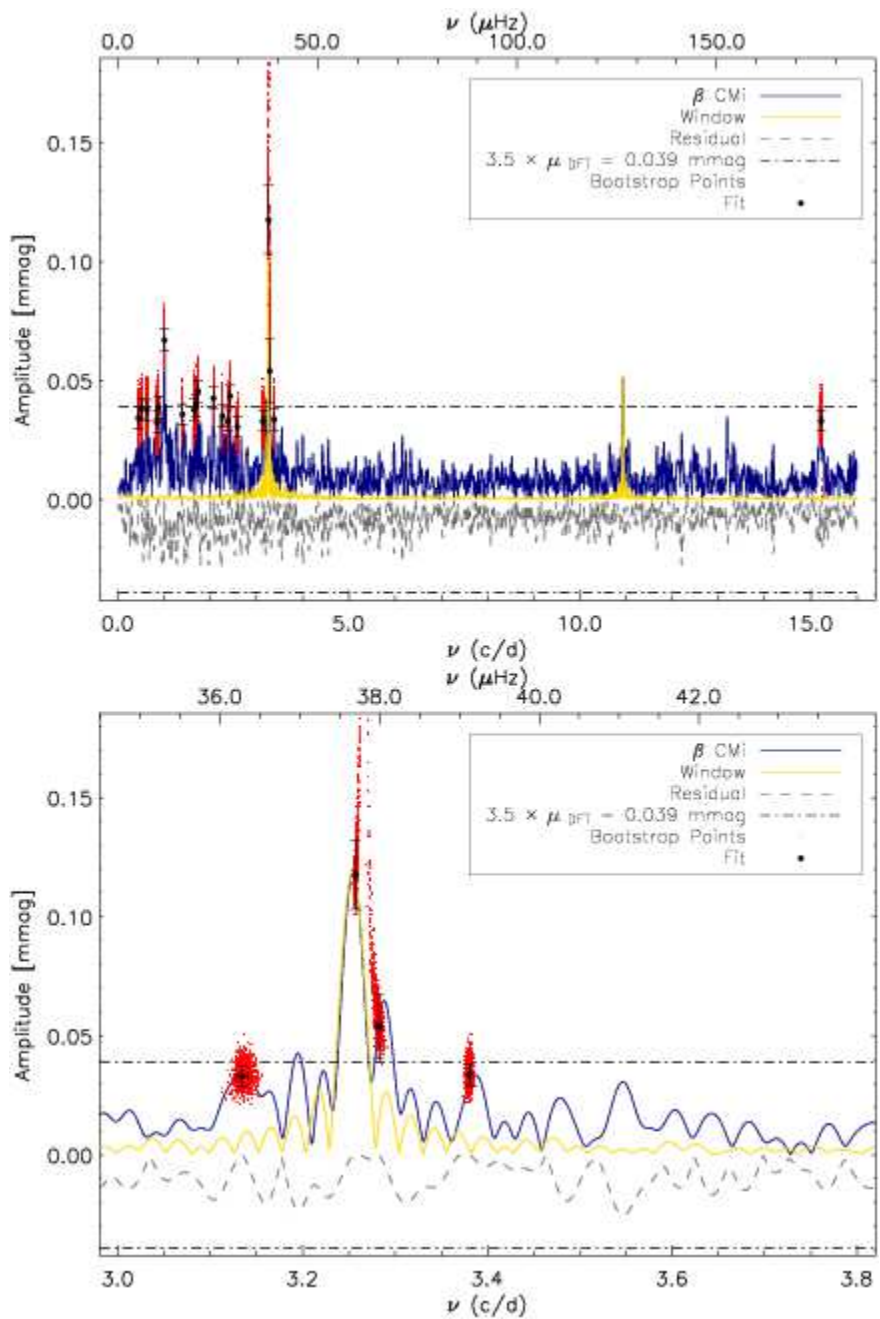}
\caption{ a) (top panel) shows the Fourier amplitude spectrum of the light curve 
in Figure 1. The points with bars are frequencies returned by CAPER having 
the largest amplitudes. The window function is shown for the strongest peak. 
Residuals after removing the 20 largest peaks are shown at the bottom of the panel
(the sign is inverted for better visibility).
Small dots indicate the frequency and amplitude obtained at each bootstrap
\citep[e.g.,][]{Wal03}, in which the frequency analysis by CAPER was 
repeated many times for data sets consisting of randomly chosen data points
\citep[see][for details]{Sai06}.
Dash-dotted lines indicate a level of 3.5 times the noise averaged over 
the spectrum from 0 to 15 cycles~day$^{-1}$.
b) (bottom panel)  shows a zoomed region around the most prominent
peak. Note that all identified frequencies are resolved.
}
\label{spectrum}
\end{figure}

Figure~\ref{spectrum} displays the discrete Fourier transform (DFT) of the MOST
light curve for $\beta~\rm{CMi}$. 
The frequency analysis was performed by CC 
using his CAPER software \citep{Wal05b,Sai06}.
Frequencies with amplitudes $\ga 0.04$ mmag 
(3.5 times the noise level averaged over the spectrum from 0 to 15 cycles~day$^{-1}$)
are identified in Figure~\ref{spectrum}.
The plot includes the spectral window function 
and shows the residual amplitude spectrum
after removal of the 20 highest peaks. The peak at 1 cycles~day$^{-1}$ is an
artifact of the Earth's rotation and the satellite's Sun-synchronous orbit,
and their modulation of parasitic light.
The peak at 15.2 cycles~day$^{-1}$ is an artifact related to {\it MOST}'s
orbital frequency.

There are two dominant features in the DFT, with frequencies of
$3.257 \pm 0.001$ cycles~day$^{-1}$ (S/N $\sim$ 8.5) and 
$3.282 \pm 0.004$ cycles~day$^{-1}$
(S/N $\sim$ 3.5). The spacing between these two frequencies is
$0.025 \pm 0.004$ cycles~day$^{-1}$.  The high S/N values for these frequencies
and detailed comparison with the background recorded in the MOST
Fabry field demonstrates that these two signals are intrinsic to
$\beta$ CMi.

There are also nearby frequencies at $3.135 \pm 0.006$ cycles~day$^{-1}$ (S/N
$\sim$ 2.9) and $3.380 \pm 0.002$ cycles~day$^{-1}$ (S/N $\sim$ 2.8).  However,
given their relatively low signal-to-noise, we do not consider
these to be definitive detections in this data set. There is also
evidence for significant stellar variability at frequencies below
3 cycles~day$^{-1}$ but its origin is unclear.

We have performed the frequency analysis on the data, both with the
long-term trends removed and left in the time series.  The frequency
identifications reported above are unaffected, which is consistent
with the very clean spectral window of the MOST photometry and the
expected lack of power leakage from low to higher frequencies in the
Fourier spectrum of these data.

Error bars for the fit parameters are determined by a bootstrap
process \citep[e.g.,][]{Wal03}. A large number of trial light curves
(10000 in this case) are generated by randomly sampling the $\beta~\rm{CMi}$ light
curve. The fitting procedure is repeated for each new light curve
resulting in a distribution for each fit parameter. 
The error bars
are 1 standard deviation in each of the parameter distributions (see
for example Saio et al 2006).
Note that these frequencies are well determined although the amplitudes
of the main peaks have moderate uncertainties.

\section{MODELS}
\begin{figure}
\epsscale{0.8}
\plotone{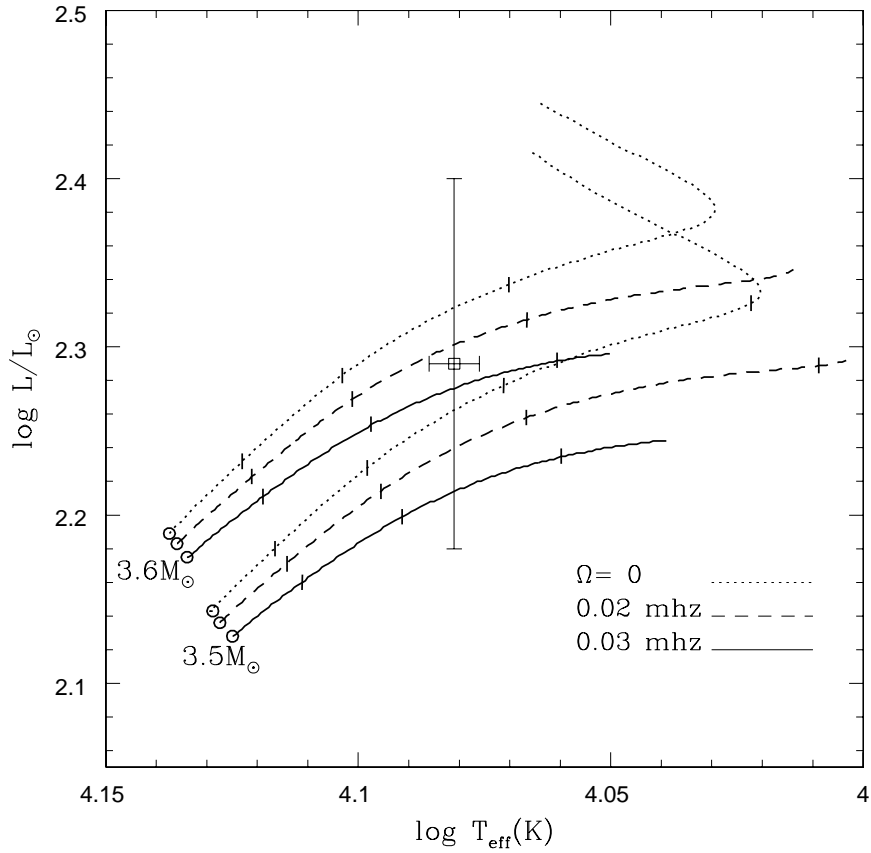}
\caption{Evolutionary tracks of $3.5M_\odot$ and $3.6M_\odot$ with
various rotation frequencies.
Circles indicate positions of zero-age-main sequence models. 
Tic marks along each evolutionary track indicate positions
separated by an interval of $5\times10^7$ yr.
The square with error bars indicates 
the approximate position of $\beta$ CMi.
}
\label{hrd}
\end{figure}

To understand the cause of  the periodic variations detected by {\it MOST} 
we have performed a pulsational stability analysis for rapidly rotating
main-sequence star models of $3.5$ and $3.6M_\odot$, whose evolutionary
tracks pass close to the position of $\beta$ CMi (Fig.\ref{hrd}). 
The evolutionary models were computed in the same way as in \citet{Wal05b},
and the stability analysis was performed based on 
the method of \citet{Lee95}.
An initial chemical
composition of $(X,Z) = (0.70, 0.02)$ is adopted. The opacity was
obtained from OPAL tables \citep{opal96}.
The centrifugal force is approximately included in the spherical 
symmetric hydrostatic equilibrium as
\begin{equation} 
{dP\over dr} = -g\rho + {2\over3}r\Omega^2,
\label{hydst}
\end{equation}
where $P$ is the pressure, $r$ is the distance from the center,
$g$ is the local gravitational acceleration, $\rho$ is the matter
density, and $\Omega$ is the angular frequency of rotation. 
The rotation frequency is assumed to be constant throughout 
the stellar interior and evolution. 
Evolutionary tracks calculated with rotation frequencies
of $0.02$ mHz (1.73 cycles~day$^{-1}$) and $0.03$ mHz (2.59 cycles~day$^{-1}$) 
are shown in Figure~\ref{hrd}. 
These tracks stop where the centrifugal force exceeds the gravitational force 
in equation (\ref{hydst}) at the stellar surface.
A track not including rotational effects has also been calculated and is
also shown in Figure~\ref{hrd}.

Figure~\ref{hrd} also shows an approximate position of $\beta$ CMi 
with error bars, which is estimated as follows.
\citet{Zor05} obtained $(\log T_{\rm eff},\log g)= (4.070,3.88)$
for $\beta$ CMi as apparent parameters and converted them to mean values
$(\log \overline{T_{\rm eff}},\log \overline{g})= (4.081,3.94)$
averaged over the whole stellar surface. These parameters yield
$\log L/L_\odot = 2.33$.  
On the other hand, \citet{Fab90} obtained an absolute magnitude of 
$M_{\rm v}=0.01$ from the data of $uvby\beta$ photometry corrected for
circumstellar emission.
Combining this value with a bolometric correction for B8V of
$-0.7$ \citep{Flo77} yields $\log L/L_\odot = 2.18$.
Furthermore, combining \citeauthor{Fab90}'s $V$ magnitude of 3.02 (corrected
for circumstellar emission) with 
the {\it Hipparcos} parallax \citep{Per97} and the above bolometric
correction, we obtain $\log L/L_\odot = 2.41$. 
From these estimates we have adopted $\log L/L_\odot = 2.29\pm 0.11$  for
the luminosity of $\beta$ CMi. For the effective temperature,
\citet{Flo77} gives $\log T_{\rm eff} = 4.086$ for the spectral type B8V.
Combining this value and \citeauthor{Zor05}'s value, we have adopted
$\log T_{\rm eff} = 4.081\pm 0.005$.

$\beta$ CMi is located close to the cool edge of
the SPB instability region of the HR diagram \citep{Pam99}. 
Our models confirm this observation (see Figure~\ref{norot}).
Along our evolutionary tracks without rotation the cooler boundary of the 
instability region appears at $\log T_{\rm eff} \approx 4.03$ 
for $l=1$, $\approx 4.06$ for $l=2$, and $\approx 4.09$ for
$l=3$. The excited modes in our $3.5M_\odot$, non-rotating models,
are high-order g-modes excited by the Fe-opacity bump 
\citep{Gau93,Dzi93}.

Since $\beta$ CMi is a rapid rotator, we have to include
the rotation effects in the stability analysis of g-modes.
Nonradial pulsations of a rapidly-rotating star
cannot be represented by a single spherical harmonic and coupled with
toroidal velocity fields. In our analysis the angular dependence
of pulsation amplitude is expanded using eight spherical harmonics 
$Y_{l_j}^m$ for a given azimuthal order $m$ ($Y_{l'_j}^m$ for toroidal
velocity field), where $l_j = |m| + 2j$ ($l'_j = l_j+1$) for even
modes and $l_j = |m| + 2j+1$ ($l'_j=l_j-1$) for odd modes with 
$j= 0, 1, \ldots, 7$
\citep[see e.g.,][for general discussions on nonradial pulsations
of rotating stars]{Unn89}. We adopt the convention that a {\it negative} 
$m$ represents a {\it prograde} mode (in the co-rotating frame) with 
respect to the stellar rotation. 
Even (odd) modes are symmetric (anti-symmetric) with respect to the
equatorial plane.  We designate the angular-dependence type of  a mode
by a set of ($m$,$\ell$), in which $\ell$ is defined as the $l_j$ value 
of the largest-amplitude component.

Figure~\ref{nute3p5} shows frequencies (in the observers' frame) 
of low surface order ($|m|\le 2$ and $\ell \le 2$) modes excited in 
evolutionary models 
of $3.5M_\odot$ rotating at frequencies of $0.02$~mHz and $0.03$~mHz
as a function of the effective temperature.
Comparing this figure with Figure~\ref{norot}, 
we find that the cooler boundary of 
instability is bluer than in the non-rotating case. 
Fewer modes are excited in more rapidly rotating models. 
This implies rapid rotation tends to stabilize g-mode 
pulsations.

The pulsation frequency in the observers' frame is given as
$$|\nu({\rm crot}) - m\nu_{\rm rot}|,$$ where  $\nu(\rm crot)$
is the pulsation frequency in the frame co-rotating at
frequency $\nu_{\rm rot}$. Therefore, pulsation frequencies
in the observer's frame for high-order g-modes and r-modes
(for which $\nu({\rm crot})\ll |m|\nu_{\rm rot}$) tend to group depending on 
the $m$ value just as seen in Figure~\ref{nute3p5}. 
All the excited prograde modes ($m < 0$) seen in Figure~\ref{nute3p5} 
are sectoral ($\ell = |m|$) g-modes, while all the excited
retrograde ($m > 0$) modes are r-modes (in which toroidal motion is
dominant) in those rapidly rotating models.
No retrograde g-modes are excited in this figure.
We note that, as shown in Appendix A, 
the stability of pulsation modes differs considerably if
we use the so-called traditional approximation in 
which the horizontal component of the angular velocity of rotation
is neglected.

Under the assumption of  $\nu_{\rm rot} \approx 0.02$ mHz 
(1.73 cycles~day$^{-1}$), 
the periodic light variations of $\beta$ CMi are 
approximately reproduced by r-modes of $(m,\ell)=(2,2)$.
On the other hand, the observed variations are identified as 
sectoral prograde g-modes of $(m,\ell)=(-1,1)$ 
if we assume $\nu_{\rm rot} \approx 0.03$ mHz (2.59 cycles~day$^{-1}$).
In the former case another three groups of frequencies are expected
to be observed in the estimated $\log T_{\rm eff}$ range of $\beta$ CMi,
while in the latter case only one group of frequencies should be 
observed (Fig.\ref{nute3p5}). 
Since {\it MOST} has detected only one group of frequencies 
around 3.26 cycles~day$^{-1}$ in $\beta$ CMi (\S\ref{most}), 
we conclude that a model rotating at  
$\nu_{\rm rot} \approx 0.03$ mHz (2.59 cycles~day$^{-1}$) is more 
appropriate.\footnote{The slower rotation model might appear to be saved 
if the low-amplitude
frequencies around 2 cycles~day$^{-1}$ in Fig.\ref{spectrum} were assigned 
to g-modes of $(m,\ell) = (-1,1)$ and the main frequencies around 
3.26 cycles~day$^{-1}$ to r-modes of $(m,\ell) = (2,2)$.
We consider this highly unlikely, however, since 
r-modes are not expected to 
yield luminosity variations much larger than the sectoral g-modes
of $(m,\ell) = (-1,1)$, and because no peaks around 4 cycles~day$^{-1}$
corresponding to g-modes of $(m,\ell) = (-2,2)$ are detected.
}
For this rotation rate pulsations excited at $\log T_{\rm eff}=4.083$ 
are prograde sectoral high-radial order (19th--33rd) 
g-modes.  The frequencies in the co-rotating frame range over
$0.586 ~{\rm cycles~day}^{-1} \ga \nu({\rm crot}) 
\ga 0.334 ~{\rm cycles~day}^{-1}$. 
Frequency spacings between two adjacent modes lie in a range from 
$0.035 $~cycles day$^{-1}$ to $0.009 $~cycles day$^{-1}$. 
Since the frequency resolution expected from the data is approximately 
$1/(41 ~{\rm days}) = 0.024$ cycles~day$^{-1}$, only a few frequencies are expected
to be resolved just as the DFT indicates (Fig.~\ref{spectrum}). 

The top panels of Figure~\ref{nute3p5} show the rotation angular frequency 
$\overline{\Omega}$ normalized by the Keplerian frequency $(GM/R^3)^{1/2}$.
For a given $\nu_{\rm rot}$, $\overline{\Omega}$ increases as the 
model evolves because the stellar radius $R$ increases.
Since the equatorial radius is larger than the mean radius, the 
rotation velocity  becomes critical at the equator with 
$\overline{\Omega}<1$; probably at $\overline{\Omega} \sim 0.8$. 
Figure~\ref{nute3p5} indicates that 
$3.5M_\odot$ models with $\nu_{\rm rot}= 0.03$ mHz rotate nearly critically 
at the effective temperature of $\beta$ CMi. 
(For $3.6M_\odot$ models with $\nu_{\rm rot}= 0.03$ mHz, 
$\overline{\Omega} > 0.8$, not appropriate for $\beta$ CMi.)
This supports Townsend et al's (2004) claim that
classical Be stars may be rotating much closer to their critical velocities 
than is generally supposed. 
It is also consistent with the recent statistical property found by 
\citet{Cra05} that the coolest Be stars rotate nearly critically.

\section{DISCUSSION}
We have found nonradial g-mode pulsations excited in the 
late-type Be star $\beta$ CMi (B8Ve). 
This confirms that $\beta$ CMi belongs to the new class of SPBe stars, like 
HD163868 \citep{Wal05b}.
We have identified pulsation modes as prograde-sectoral g-modes of $m=-1$,
based on the comparison between the observed frequencies and theoretical  
ones excited by the Fe bump of opacity in rapidly rotating models.
The fact that only one group of frequencies is detected by {\it MOST} is consistent
with our prediction of the pulsational stability 
if $\beta$ CMi rotates nearly 
critically. This supports the recent claims \citep{Tow04,Cra05} that 
classical Be stars rotate more rapidly than previously thought  
\citep[70--80\% of the critical rate; e.g.][]{Por96,Cha01}.
The observed amplitudes are small ($\la 1$ mmag). 
Since $\beta$ CMi lies close to the red edge of the instability
region on the HR diagram, the pulsational instability is not
strong and hence the pulsation amplitudes are expected to be small.

Our discovery of nonradial pulsations in $\beta$ CMi suggests the possibility 
that nonradial pulsations are involved in all rapidly rotating Be stars.
In some early-type Be stars mass ejections are observed to occur
at certain pulsation phases \citep[e.g. $\mu$ Cen:][]{Por03} suggesting
that pulsations trigger mass ejections.
It is tempting to speculate that nonradial pulsations may play a 
crucial role in mass ejection in all Be stars.
\citet{Osa86} argued that prograde nonradial pulsations can 
transport angular momentum toward the surface and spin up the 
equatorial region critically leading to a mass ejection. 
\citet{Owo05} showed numerically that introducing velocity variations
corresponding to a prograde ($m=-2$) nonradial pulsation in a nearly
critically rotating stellar atmosphere actually yields a circumstellar 
Keplerian disk.

Our findings in this paper and \citet{Wal05b}, that 
{\it prograde} sectoral g-modes are selectively excited in rapidly 
rotating Be stars, support the above scenario for mass ejection.
Line-profile-variation (LPV) analyses for early type Be stars, 
however, tend to 
indicate the presence of retrograde modes \citep{Riv03} rather than prograde
modes, contrary to our model findings.

\citet{Riv98} obtained four periods around 0.503 days 
and two period around 0.28 days 
from LPVs in $\mu$ Cen, one of the best studied and most active Be stars.
[These LPVs of $\mu$ Cen were modeled as retrograde
($m=2,~{\rm and}~3$) modes by \citet{Riv01}.] 
Frequency spacings among each group are 
about 0.034 cycles~day$^{-1}$ or less,
which are comparable with the frequency difference 0.025 cycles~day$^{-1}$ 
between the two main frequencies of $\beta$ CMi (\S2) and with
typical frequency spacings $\sim 0.03$ cycles~day$^{-1}$ of HD 163868
\citep{Wal05b}. These frequency spacings are comparable
with those of g-modes, although they are close to the limit
of the frequency resolution. 
Longer observations by MOST or COROT would make g-mode
seismology of SPBe stars possible.
Also, simultaneous photometric and spectroscopic 
observations of $\mu$ Cen and other Be stars
showing conspicuous LPVs and mass-ejection events would be important
to understand the connection between nonradial pulsations and mass ejections
in Be stars.

\acknowledgments 
We thank Thomas Rivinius, the referee of this paper, for his 
thoughtful comments.
This research has made use of the SIMBAD
database, operated at CDS, Strasbourg, France. The Natural Sciences
and Engineering Research Council of Canada supports the research of C.C.,
D.B.G., J.M.M., A.F.J.M., J.F.R., S.M.R., G.A.H.W.  A.F.J.M. is also
supported by FQRNT (Qu\'ebec). R.K. is supported by the Canadian Space
Agency. W.W.W. is supported by the Austrian Space Agency and the
Austrian Science Fund (P17580-N2).  H.S. is supported by the 21st Century
COE programme of MEXT, Japan.

\appendix

\section{Mode stability under the traditional approximation}
The pulsation analysis for low-frequency modes of a rapidly rotating star
is greatly simplified by the use of the so-called traditional approximation,
in which the horizontal component of the angular velocity of rotation
($\Omega\sin\theta$ with $\theta$ being the co-latitude) is neglected.
Under the traditional approximation, eigenfrequencies can be obtained
using the same equations as those for nonradial pulsations 
of non-rotating stars
except replacing $\ell(\ell+1)$ with the parameter $\lambda_{\ell,m}$
\citep[see e.g.,][]{Tow05,Lee06}.  
The value of $\lambda_{\ell,m}$ deviates from $\ell(\ell+1)$
as the ratio of the rotation to the pulsation frequency 
(in the co-rotating frame) 
increases depending on $m$ and $\ell$ \citep[e.g.,][]{Lee97}. 

\begin{figure}
\epsscale{1.0}
\plotone{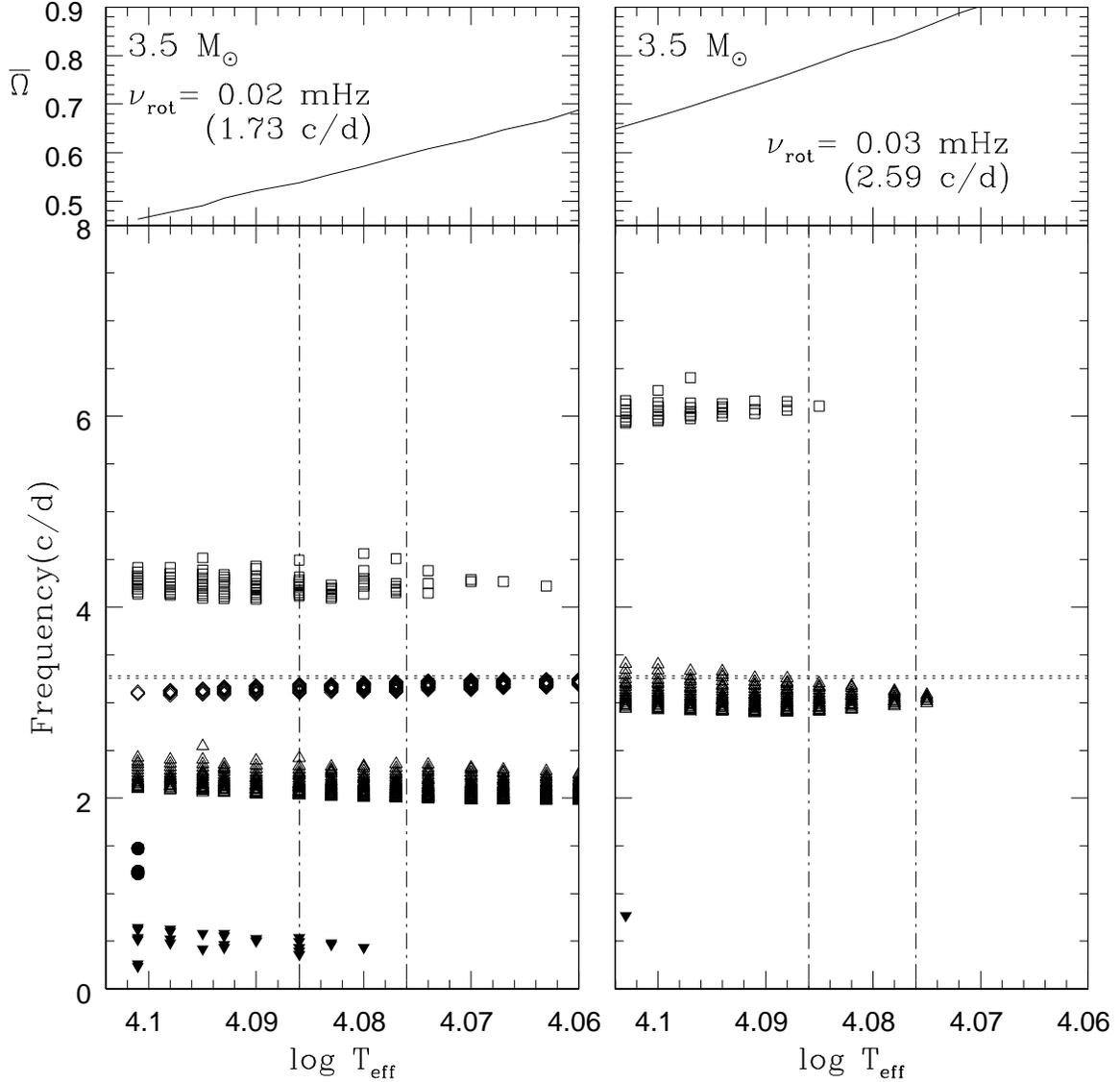}
\caption{
Frequencies in the observer's frame of excited pulsation modes 
(bottom panels) and the angular velocity of rotation normalized by $(GM/R^3)^{1/2}$
(top panels) as functions of effective temperature.
Horizontal dotted lines indicate observed frequencies
3.257 cycles~day$^{-1}$ and 3.282 cycles day$^{-1}$ for $\beta$ CMi.
Vertical dash-dotted lines indicate the estimated range for 
the effective temperature of $\beta$ CMi. 
Rotational frequencies are assumed to be 0.02 mHz (left-hand panel)
and 0.03 mHz (right-hand panel) through the main-sequence evolution.
Symbols indicate type of modes: 
filled circles ($\bullet$) for $m = 0, \ell = 1$;
triangles ($\triangle$) for $m = -1, \ell = 1$;
squares ($\Box$) for $m=-2, \ell = 2$;
inverted triangles ($\blacktriangledown$) for $m = 1, \ell = 2$;
diamond ($\diamond$) for $m = 2, \ell = 2$.
Only r-modes are excited among retrograde modes ($m > 0$).
Other excited modes are sectoral prograde g-modes. (A few axisymmetric
g-modes are also excited in the hottest models).
}
\label{nute3p5}
\end{figure}

\begin{figure}
\epsscale{0.6}
\plotone{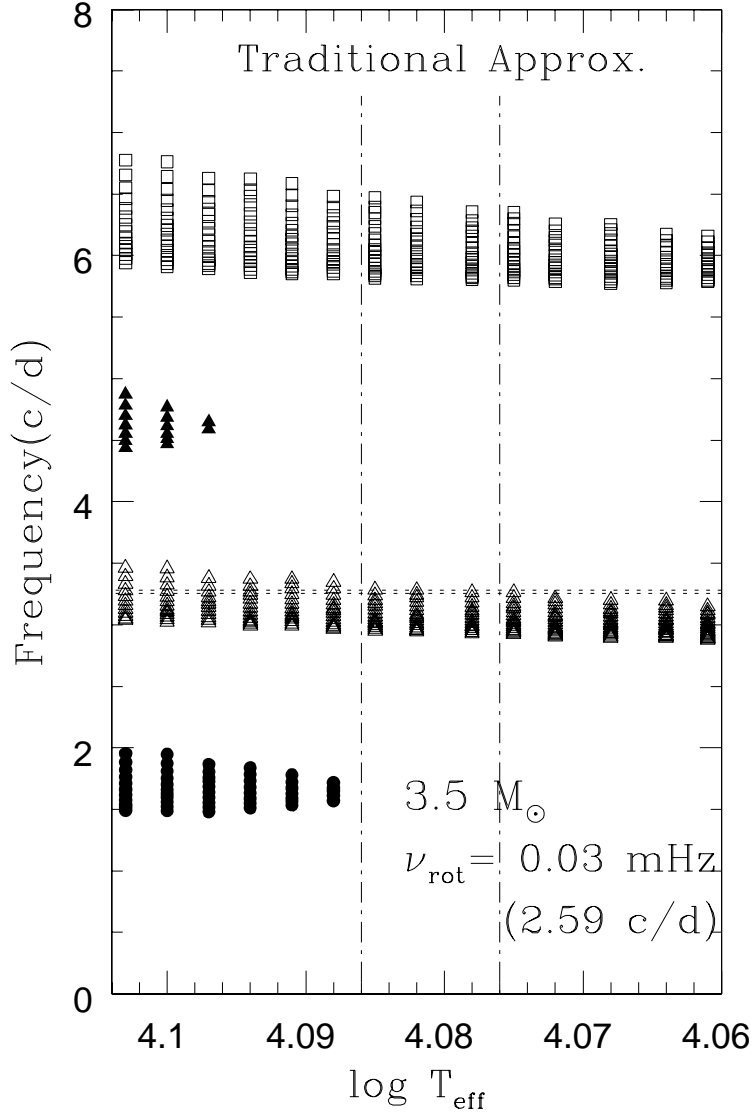}
\caption{
The same as the right-bottom panel of Figure~\ref{nute3p5} but showing
results obtained under the traditional approximation.
Filled triangles ($\blacktriangle$) are for g-modes of 
$(m,\ell)=(-1,2)$, and other symbols are the same as in Figure~\ref{nute3p5}.
Compared to Figure~\ref{nute3p5} more sectoral g-modes of $m=-1$, and $-2$ are
excited.  Tesseral g-modes of $(m,\ell)=(0,1)$ (around 1.7 cycles~day$^{-1}$) and
of $(m,\ell)=(-1,2)$ (around 4.7 cycles~day$^{-1}$) are excited only under
the traditional approximation in these models. 
}
\label{trad_nute}
\end{figure}

The traditional approximation is justified dynamically 
for low-frequency pulsations, but it is not clear whether 
the approximation is reasonable for the nonadiabatic stability analysis.
To see the effect, we have performed a stability analysis using the traditional
approximation for $3.5$-$M_\odot$ 
models rotating at a frequency of 0.03 mHz. 
The results are shown in Figure~\ref{trad_nute}.
Comparing this figure with Figure~\ref{nute3p5}, we see that 
under the traditional approximation, larger numbers of prograde sectoral 
g-modes are excited, and that
tesseral g-modes of $(m,\ell) = (0, 1)$ and $(-1,2)$ are also excited. 
These tesseral modes are all 
damped without the traditional approximation in these models.
The cause of the difference is attributed to damping due to mode-couplings
\citep{Lee01} which are absent under the traditional approximation.
The damping effect is strongest for retrograde g-modes 
and tesseral ($\ell > |m|$) g-modes.

\begin{figure}
\epsscale{1.0}
\plotone{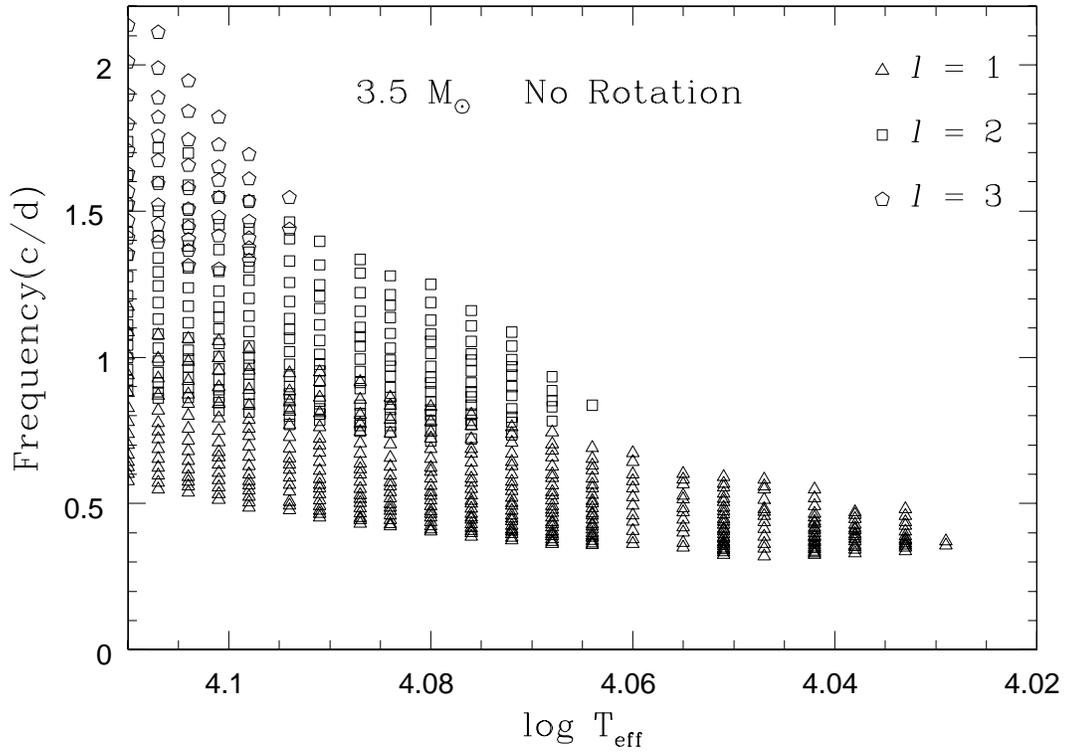}
\caption{
Frequencies of g-modes excited
in non-rotating models along the $3.5M_\odot$ main-sequence evolution.
The evolutionary track is shown in Fig.~\ref{hrd}.
}
\label{norot}
\end{figure}

We note that in the models shown in Figure~\ref{trad_nute} no retrograde g-modes
are excited even under the traditional approximation.
This is understood as follows. Retrograde g-modes have large values of 
$\lambda_{\ell,m}$ \citep{Lee97}, which correspond to modes of large 
latitudinal degree $\ell$ in a non-rotating model. 
Figure~\ref{norot} for nonrotating models indicates 
that the cooler boundary of instability region
is hotter for larger $\ell$. Therefore, we understand that the temperature
range for instability of retrograde g-modes is hotter than that
shown in Figure~\ref{trad_nute}.

\clearpage

\clearpage

\clearpage

\clearpage

\clearpage

\clearpage

\end{document}